\documentclass[12pt]{article}

\usepackage[margin=1in]{geometry}
\usepackage{amsmath}
\usepackage{amssymb}
\usepackage{array}
\usepackage{authblk}

\usepackage{booktabs}
\usepackage{float}
\usepackage{graphicx}
\usepackage{algorithm}
\usepackage{algpseudocode}
\usepackage{hyperref}
\usepackage{microtype}
\usepackage{natbib}
\usepackage{placeins}
\usepackage{url}

\graphicspath{{figures/}}
\newcolumntype{P}[1]{>{\raggedright\arraybackslash}p{#1}}
\algrenewcommand\algorithmicrequire{\textbf{Input:}}
\algrenewcommand\algorithmicensure{\textbf{Output:}}

\title{Travel Mode- and Purpose-Specific Origin–Destination Matrices for England and Wales from Fused Travel Survey and Mobile Network Data}
\author[1]{Bowen Zhang}
\author[1]{Chen Zhong}
\author[2]{Mingfei Ma}
\author[2]{James Golding-Graham}
\affil[1]{The Bartlett Centre for Advanced Spatial Analysis, University College London, Gower Street, London, WC1E 6BT, UK}
\affil[2]{England's Economic Heartland business unit, c/o Buckinghamshire Council, Walton Street, Aylesbury, HP20 1UA, UK}
\date{}

\begin{document}
\maketitle

\begin{abstract}
Origin--destination (OD) matrices sit behind much of quantitative transport planning, from model calibration and accessibility analysis to the appraisal of new services and development. The increasing emphasis on place-based solutions requires mobility data that can support decision-making not only at the strategic level, but also at finer spatial scales. This requires up-to-date OD evidence at small-area resolution, disaggregated by travel mode and purpose, which remains either inaccessible or unavailable. In this work, we present dense MSOA-to-MSOA OD matrices for England and Wales, segmented by seven travel modes and representative time periods, with eight trip purposes for the weekday morning peak. The matrices are built by calibrating aggregate mobile network data provided by BT against National Travel Survey (NTS), census and trip-rate evidence, preserving the observed spatial structure of movement while referencing its age, mode and purpose composition to the survey. The open-source data processing pipeline is released alongside the matrices, so that the construction of the dataset can be inspected in full and adapted to other years, regions or assumptions.
\end{abstract}


\section{Introduction}

Origin--destination (OD) matrices record how many trips run between each pair of places, and they sit behind most quantitative work on how a city or region moves: transport model calibration, accessibility and catchment measurement, spatial interaction modelling, and the business case for a new bus route or housing allocation, where the argument turns on an estimate of who would travel where as a result. That work needs OD evidence at small-area resolution, since the decisions themselves are local. This expectation is now written into policy. The statutory guidance for local transport plans in England asks authorities to make place-based assessments of how people move within \emph{and beyond} their administrative area, observing that travel-to-work areas frequently do not align with authority boundaries \citep{dft2026localTransportPlans}. In addition, that work needs evidence segmented by mode and purpose: for example, an intervention aimed at the school run is not the same as one aimed at peak commuting. Analysts rarely get both at once, and must choose instead among sources that are strong in different places, a division that has kept survey-based and passively-collected mobility research largely separate \citep{chen2016promisesBigSmallData}. 

The most relevant datasets in the UK, including Census journey-to-work flows, are open and spatially detailed, but they cover a single trip purpose and are collected once a decade \citep{ons2023originDestination}. The National Travel Survey (NTS) records mode, purpose and time of day in depth, but it is a household survey: the 2024 round achieved 8,975 participating households, all of them in England \citep{dft2025ntsTechnicalReport}. Spread across the 7,264 MSOAs of England and Wales that is roughly one household per zone, so the survey cannot support small-area OD estimation \citep{dft2024nationalTravelSurvey}. Passive mobile network data observe movement at population scale, and a substantial literature shows they can support OD and travel-demand estimation \citep{calabrese2011estimatingOriginDestination,iqbal2014developmentOriginDestination,alexander2015originDestinationTrips,toole2015pathMostTraveled}. These are aggregate network products rather than travel surveys, so their coverage follows what a mobile network observes: trips are counted for device-carrying adults, mode is inferred from the movement trace rather than reported by the traveller, and trip purpose lies outside what the source records. They are also licensed for defined uses, so a product derived from them cannot simply be passed on.

This paper describes a fused dataset that integrates mobile and survey sources and produces dense MSOA-to-MSOA OD matrices for England and Wales, split into seven travel modes and representative time periods, with eight trip purposes for the weekday morning peak. The mobile records contribute the observed spatial structure of movement, at a resolution no survey sample can reach. The NTS, census populations and trip-production rates then add the dimensions those aggregates do not carry: travel by children, a survey-referenced mode mix, and trip purpose. The fusion runs as a staged calibration, and every stage is implemented in the published pipeline, so the evidence and assumptions behind any published value can be followed in the code that produced it. To the best of our knowledge, it is the first openly released OD product for England and Wales that combines national small-area coverage with mode, purpose and time segmentation. The pipeline that builds it, \texttt{uk-travel-pipeline}, is published as supporting material so the matrices can be inspected, reproduced and rerun for other years or study areas.

\section{Data and code release}
\label{sec:product}

The primary files are $N \times N$ OD matrices in CSV format at MSOA level for England and Wales (7,264 MSOAs), each cell reporting a trip volume. Matrices are provided for two temporal groupings --- a representative week and a weekday AM peak --- and for seven travel modes: walking, cycle, private car, motorcycle, bus, rail and subway.

The matrices describe a 56-week observation window running from September 2024 to September 2025, made up of 280 weekdays and 112 weekend days, of which 229 and 97 respectively carried usable observations. The two groupings summarise that window in different ways. For the representative week, each record's volume is divided by the number of days on which it was observed and multiplied by five for weekdays or two for weekend days, then summed over the three daily periods the operator reports; a cell is therefore the trips expected in an average seven-day week. Those periods are assigned by departure time from the origin, and cover 07:00--09:59 for the AM peak, 16:00--18:59 for the PM peak, and all remaining hours for off-peak. The weekday AM peak grouping instead reports an average \emph{single} weekday 07:00--09:59 period, not a weekly total. Calibration targets come from the National Travel Survey: mode shares from the 2024 survey year of table NTS9916 \citep{dftNts9916}, and weekday purpose shares from the 2023--24 survey year of table NTS0502 \citep{dftNts0502}. Both cover England only.

The two groupings are released at different processing stages (Table~\ref{tab:products}): the representative week as one all-purpose matrix per mode, and the weekday AM peak as one all-purpose matrix per mode plus one matrix per mode and purpose. Purpose is released for the AM peak only because the trip-production rates behind the purpose allocation are counted per activity rather than per trip: an outbound journey and its return form a single activity, and the rates do not record when the return is made, so they cannot be aligned directly with the BT records, each of which is a single directional trip. In the morning peak most trips are outbound, which keeps this mismatch small, although it does not remove it. Summing the eight purpose matrices for a mode recovers that mode's all-purpose AM-peak matrix, and the release holds $7 + 7 + 7 \times 8 = 70$ matrix files in total. Alongside the matrices the release carries the \texttt{uk-travel-pipeline} source and CLI, and documentation; release metadata map purpose codes to descriptions and record which pipeline options produced the release.

\begin{table}[htbp]
\centering
\scriptsize
\caption{Released products and the processing stage at which each is taken.}
\label{tab:products}
\begin{tabular}{@{}P{0.15\textwidth}P{0.24\textwidth}P{0.21\textwidth}P{0.30\textwidth}@{}}
\toprule
Filename prefix & Contents & Unit of each cell & Processing stage \\
\midrule
\path{typical_week_} & Seven all-purpose matrices, one per mode, covering all daily periods & Estimated person trips per representative seven-day week & After child-origin uplift, mode-share calibration and road-mode split; before mode--time refinement and purpose allocation \\
\path{weekday_AMpeak_} & Seven all-purpose matrices, and 56 purpose-specific matrices (seven modes $\times$ eight purposes) & Estimated person trips per average single weekday, 07:00--09:59 & After mode--time refinement; purpose-specific files also after purpose allocation and raking to NTS weekday purpose shares \\
\bottomrule
\end{tabular}
\end{table}

Dimensions are encoded in the folder and file names rather than in columns; Table~\ref{tab:schema} gives the equivalent tidy schema used in the metadata and catalogue files, and applicable to optional long-form exports.

\begin{table}[htbp]
\centering
\small
\caption{Equivalent long-form schema for the published matrix files.}
\label{tab:schema}
\begin{tabular}{p{0.22\textwidth}p{0.14\textwidth}p{0.52\textwidth}}
\toprule
Column name & Data type & Description \\
\midrule
Origin & String & Origin MSOA 2021 code (MSOA21CD). \\
Destination & String & Destination MSOA 2021 code (MSOA21CD). \\
Travel Mode & String & One of walking, cycle, private car, motorcycle, bus, rail, subway; encoded in the filename. \\
Travel Purpose & String & Estimated purpose code 1--8, weekday AM-peak purpose files only; encoded in the filename and mapped in release metadata. \\
Period & String & Output grouping: \path{typical_week} (average seven-day week) or \path{weekday_AMpeak} (average single weekday 07:00--09:59 period); encoded in the folder name. \\
Volume & Float & Synthetic trip volume for the September 2024--September 2025 reference window, at the processing stage given for its product in Table~\ref{tab:products}. \\
\bottomrule
\end{tabular}
\end{table}

Each calibration step also writes a diagnostic file recording what it changed and by how much: child-uplift factors for all 7,264 origins, the mode-share check (452 rows, one for each combination of origin region, NTS trip-length band and output mode, 427 of them with an NTS target) and its 275 bounded adjustment factors, the mode--time share check and its factors, and the NTS0502 purpose control check. These are quality-assurance outputs retained for internal review and are not distributed with the release, but they are what the checks in Section~\ref{sec:checks} are computed from, and the pipeline regenerates them on every run, so any user rebuilding the dataset obtains the equivalent record for their own build.

\section{Methods: Dataset construction and fusion}
\label{sec:construction}

Five input streams feed the build (Table~\ref{tab:inputs}). Only the BT mobile network aggregate is commercially licensed; the other four are open or available to researchers under standard licences \citep{ons2021msoaBoundaries,dft2024nationalTravelSurvey,dft2024nationalTravelSurveySpecialLicence,tfnTripRates,tfnLandUse}. Because only the mobile base is licensed, the pipeline can be rerun outside the original project by any user holding a licence for a compatible feed.

\begin{table}[htbp]
\centering
\scriptsize
\caption{Input datasets and their role in the workflow.}
\label{tab:inputs}
\begin{tabular}{@{}P{0.26\textwidth}P{0.30\textwidth}P{0.33\textwidth}@{}}
\toprule
Data source & Content & Role in the pipeline \\
\midrule
BT mobile network aggregate modal-share data & Origin and destination MSOA, inferred mode, period, weekend flag, adult volume, days observed. & Observed adult OD pattern and initial modal assignment. Commercially licensed; not redistributed. \\
ONS MSOA 2021 boundaries and lookup tables \citep{ons2021msoaBoundaries} & MSOA codes, centroids, geometries, NTS region identifiers. & Standardises records to MSOA 2021 geography; assigns OD distances, trip-length bands and origin regions. \\
NTS9916 regional trip-length evidence \citep{dftNts9916} & Trips by year, region of residence, trip-length band and mode. & Target mode shares for the region--band calibration. \\
Census-derived and TfN NorMITs Land Use population segment data \citep{tfnLandUse} & LSOA/MSOA population by demographic, household and area-type segment. & Child-origin uplift and local weighting of trip-production priors. \\
NTS/TfN trip-production rate tables \citep{tfnTripRates}; NTS0502 period--purpose shares \citep{dftNts0502} & Trip rates by mode, period, purpose, household and area type; weekday period--purpose control shares. & Allocates adjusted volumes to purposes; constrains origin--period mode totals. \\
\bottomrule
\end{tabular}
\end{table}

The raw mobile input is not redistributed; deposit locations, licences and disclosure constraints for the published product are given in the data availability statement.

\subsection{Data fusion and calibration pipeline}

The pipeline runs four sequential steps, each adding a dimension the mobile aggregates do not carry or referencing one to survey evidence, and each switchable so a sensitivity build differs from the main build only in configuration. A schematic of the whole workflow, showing which input feeds which step, is given in the supplementary material. Throughout, $V_{ijmt}$ is the trip volume from origin MSOA $i$ to destination MSOA $j$ by mode $m$ in time period $t$, and a superscript marks how far that volume has been processed: $V^{\mathit{adult}}$ as delivered, $V^{\mathit{base}}$ after the all-age uplift, and $V^{*}$ after mode calibration. Calibration operates on groups of records rather than on single records, so each record also carries the origin region $r$ that contains $i$ and the trip-length band $b$ into which its OD distance falls.

\subsubsection*{Step 1: Standardising geography and building an all-age base}

A BT record is a directional flow between one origin and one destination MSOA for a single inferred mode, time period and weekday/weekend flag, carrying an adult trip volume and the number of days over which it was observed. Records are aggregated to the MSOA 2021 boundary and joined to zone centroids and the region-of-residence lookup \citep{ons2021msoaBoundaries}. Straight-line distances between origin and destination centroids are converted to miles and assigned to the eight NTS trip-length bands: under 1 mile, 1 to under 2, 2 to under 5, 5 to under 10, 10 to under 25, 25 to under 50, 50 to under 100, and 100 miles and over. Every record therefore carries the $(r,b)$ keys that the next step calibrates on.

Because the BT mobile data are scoped to adults, volumes at each origin are then uplifted to all ages by the local population ratio
\[
U_i = \frac{\mathit{AdultPop}_i + \mathit{ChildPop}_i}{\mathit{AdultPop}_i},
\qquad
V^{\mathit{base}}_{ijmt} = V^{\mathit{adult}}_{ijmt}\, U_i,
\]
where $\mathit{AdultPop}_i$ and $\mathit{ChildPop}_i$ are the adult and child populations resident in origin $i$. The mean uplift in the national run is 1.226, corresponding to a mean origin child share of 18.3\%. Added child travel inherits the adult OD pattern of its origin, a limitation we return to in Section~\ref{sec:limitations}.

\subsubsection*{Step 2: Mode-share calibration}

Mode calibration is the core of the pipeline, and the one step where the mode mix of the dataset is set by the survey rather than by the mobile source. Table NTS9916 \citep{dftNts9916} reports the average number of trips by mode for each region of residence and trip-length band, from the 2024 survey year. Its thirteen survey modes are collapsed onto the four broad classes the BT feed distinguishes --- walking, road, rail and subway --- giving a target mode share for each of the nine English regions of residence and each of the eight trip-length bands above. Because the NTS has covered England only since 2013 \citep{dft2025ntsTechnicalReport}, the table provides no Welsh targets. Within each region--band group, the share of volume travelling by mode $m$ is
\[
S^{\mathit{base}}_{rbm} = \frac{\sum_{ijt \in (r,b)} V^{\mathit{base}}_{ijmt}}{\sum_{ijt \in (r,b)} \sum_{m'} V^{\mathit{base}}_{ijm't}},
\]
that is, the volume moving by mode $m$ divided by the volume moving by any mode $m'$ in the same region and band. Comparing this observed share with the corresponding NTS target share $S^{\mathit{NTS}}_{rbm}$ gives an adjustment factor, which is then applied to every individual record in that group:
\[
A_{rbm} = \frac{S_{rbm}^{\mathit{NTS}}}{S_{rbm}^{\mathit{base}} + \epsilon},
\qquad
V_{ijmt}^{*} = V_{ijmt}^{\mathit{base}}\, A_{rbm},
\]
where $\epsilon = 10^{-12}$ guards against division by zero where a mode is unobserved. Factors are clipped to $[0.01, 100]$ by default, so that a mode seen in only a handful of records cannot be scaled by an implausible amount; groups for which the survey gives no target keep $A_{rbm} = 1$, and the median factor in the national run is 0.968. In the national build these untargeted groups are exactly the Welsh ones: 25 of the 275 region--band--mode groups, all of them in Wales, receive no mode-share correction and retain the mode composition of the uplifted mobile base.

A single pass suffices, because each factor is computed from exactly the totals it is then applied to: after application the group shares equal the NTS targets, up to the effect of the clipping bounds. The step rescales volumes within a region--band--mode group but never moves a trip from one OD pair to another, so the spatial structure observed by the mobile data is preserved while its mode composition is referenced to the survey.

\subsubsection*{Step 3: Road-mode split and mode--time refinement}

Two optional refinements make the mode dimension planning-ready. The original BT data classify all motor-based modes into one broad class, \texttt{ROAD}. We split this into cycle, private car, motorcycle and bus using NTS-derived shares for the record's region and trip-length band; the shares sum to one within each cell, so the split conserves the parent volume, and cells without NTS road evidence default to private car. This yields the seven output modes: walking, cycle, private car, motorcycle, bus, rail and subway.

The second refinement corrects the time-of-day profile of each mode. Mobile observation is uneven across the day, so the share of an origin's travel that a given mode takes in the morning peak need not match survey evidence. Let $q$ index mode--time groups --- the seven output modes, with motorcycle constrained jointly with private car because the trip-rate tables report them together --- and let $V_{itq}$ be the volume assigned to group $q$ at origin $i$ in period $t$ after Step~2. Writing $N_{it} = \sum_q V_{itq}$ for the total already assigned there, and $\sigma_{itq}$ for the share that the NTS/TfN mode--time split table \citep{tfnTripRates} gives to that group, the target shares are first renormalised over the groups actually observed at that origin and period and then applied to the existing total:
\[
\tilde{\sigma}_{itq} = \frac{\sigma_{itq}}{\sum_{q' :\, V_{itq'} > 0} \sigma_{itq'}},
\qquad
V'_{itq} = N_{it}\, \tilde{\sigma}_{itq}.
\]
Renormalising is what keeps the correction honest. Since $\sum_q \tilde{\sigma}_{itq} = 1$, the origin--period total $N_{it}$ is conserved exactly; and restricting that sum to groups with observed volume means no travel is invented for a mode the mobile data never recorded at that origin. Origin--period cells for which the tables give no target retain their observed shares. The resulting per-record factors $V'_{itq} / V_{itq}$ are folded into the cumulative adjustment factor, and both refinements write their own share-check and factor files.

\subsubsection*{Step 4: Purpose allocation and raking to survey controls}

Purpose is not recorded in the mobile input at all, and is estimated entirely from survey and census evidence. The estimate rests on \emph{trip-production rates}: the average number of activities a person in a given population segment undertakes for a given purpose, by mode and time period, published in the NTS/TfN trip-rate tables \citep{tfnTripRates}, where an outbound journey and its return count as one activity. Multiplying those rates by the population actually resident in a zone gives the activities that zone would be expected to generate for each purpose. Concretely, for each origin the resident population segments --- each segment $c$ being a combination of TfN area type and household type --- are weighted by their trip-production rates to give a local prior over purposes:
\[
W_{imtp} = \sum_{c} \mathit{Pop}_{ic}\, \mathit{Rate}_{cmtp},
\qquad
P_{imtp} = \frac{W_{imtp}}{\sum_{p'} W_{imtp'}},
\qquad
Y_{ijmtp} = V^{*}_{ijmt}\, P_{imtp},
\]

where $\mathit{Pop}_{ic}$ is the population of segment $c$ living in origin $i$, and $\mathit{Rate}_{cmtp}$ the rate at which that segment makes purpose-$p$ trips by mode $m$ in period $t$. Their product summed over segments, $W_{imtp}$, is the expected activity count for the origin; normalising it across purposes gives the shares $P_{imtp}$, which split the calibrated volume into $Y_{ijmtp}$. The proportion of trips assigned to different purposes is not the same everywhere, because the weights come from each origin's own demographic mix and purpose composition varies from place to place.

The local prior decides how one origin's trips divide between purposes, but it says nothing about the national picture: summed over the whole country, the purpose mix is whatever the trip-rate tables happen to imply. An optional final stage anchors that mix to published survey controls, while leaving each origin free to differ from the national average according to who lives there. The control table, NTS0502 \citep{dftNts0502}, reports trip start time by trip purpose for England and covers Monday to Friday only; we use the 2023--24 survey year. Its weekday-only scope is why the raking below controls weekday periods and leaves weekend travel to the local prior.

It helps to picture the allocation as a table: one row for each combination of origin, mode and weekday period, and one column for each trip purpose. Every cell holds the volume currently assigned to that purpose, and raking adjusts those cells until two conditions hold at once.

Reading along a row, the purpose volumes must still sum to the volume already calibrated for that origin, mode and period, $R_{imt} = \sum_j V^{*}_{ijmt}$, because splitting trips between purposes must not change how many trips there are. Reading down the columns, each purpose control group $g$ --- a single purpose, or the pair that NTS0502 reports jointly --- must take its published share $T_{tg}$ of that period's trips.

The two conditions pull against each other. Scaling cells so that a control group reaches its target $T_{tg}$ throws the row sums off $R_{imt}$; rescaling each row back to $R_{imt}$ then shifts the control-group totals away from $T_{tg}$ again. Neither correction can be applied once and left alone. Iterative proportional fitting resolves this by applying them in alternation: each pass leaves a smaller discrepancy than the last, and the loop stops once both conditions hold to a set tolerance (Algorithm~\ref{alg:rake}).

\begin{algorithm}[H]
\caption{Raking purpose allocations to NTS0502 controls.}
\label{alg:rake}
\begin{algorithmic}[1]
\Require Local-prior shares $P_{imtp}$; row totals $R_{imt} = \sum_j V^{*}_{ijmt}$; control shares $s^{0502}_{tg}$; tolerance $\tau = 10^{-6}$; iteration cap $K = 100$
\Ensure Purpose shares meeting both the row and control constraints
\State $Y_{imtp} \gets R_{imt}\, P_{imtp}$ \Comment{start from the local prior}
\State $T_{tg} \gets s^{0502}_{tg} \sum_{i,m} R_{imt}$ \Comment{control target per group and weekday period}
\State $C \gets \{(i,m,t,p) : \text{purpose } p \text{ is controlled in period } t\}$
\ForAll{control groups $g$, weekday periods $t$ with $T_{tg} > \tau$}
  \If{$\sum_{(i,m,p)\in C,\, g(p)=g} Y_{imtp} = 0$} \textbf{abort} \Comment{no local prior supports a controlled purpose} \EndIf
\EndFor
\For{$k \gets 1$ \textbf{to} $K$}
  \ForAll{$(i,m,t,p) \in C$} \Comment{control update: controlled cells only}
    \State $\sigma \gets \sum_{(i',m',p')\in C,\, g(p')=g(p)} Y_{i'm'tp'}$
    \State $Y_{imtp} \gets Y_{imtp}\cdot(\,\sigma>0 \mathrel{?} T_{t\,g(p)}/\sigma : 1\,)$
  \EndFor
  \ForAll{rows $(i,m,t)$} \Comment{row update: every row, controlled or not}
    \State $\rho \gets \sum_{p'} Y_{imtp'}$;\quad $Y_{imtp} \gets Y_{imtp}\cdot(\,\rho>0 \mathrel{?} R_{imt}/\rho : 1\,)$ for all $p$
  \EndFor
  \State $\delta \gets \max\big(\max_{t,g}|\textstyle\sum_{g(p)=g}Y_{imtp}-T_{tg}|,\ \max_{i,m,t}|\sum_p Y_{imtp}-R_{imt}|\big)$
  \State \textbf{if} $\delta \le \tau$ \textbf{then break}
\EndFor
\State \Return $Y_{imtp}/R_{imt}$ where $R_{imt}>0$, else $0$
\end{algorithmic}
\end{algorithm}

The two updates act on different cells: the control update rescales only those rows that carry an NTS0502 target, while the row update rescales every row. Each is guarded, so a group or row that is momentarily empty is left unchanged rather than dividing by zero. Convergence is tested after both updates, so the loop always completes at least one full pass and the residual is measured after the row update; because the row update comes last, purpose shares sum to one exactly within every row, and any remaining discrepancy falls on the control totals, which the national build closes to within 0.01 percentage points. Before iterating, the pipeline aborts if any controlled purpose has no local-prior support, rather than inventing trips the local population does not generate. Control groups follow NTS0502 reporting: social/leisure and visiting friends and family are controlled together and split within the group by the local prior, while the other purposes are controlled separately. Weekend periods are not controlled and keep their local-prior shares. The converged shares depend on origin, mode and period but not destination, so raking, like the mode calibration, never moves trips between OD pairs.

\subsection{Implementation}

The implementation is a plain Python package, \texttt{uk-travel-pipeline} 0.3.0, built on pandas, Dask, GeoPandas, NumPy and PyArrow \citep{pandasDevelopmentTeam2024pandas,daskDevelopmentTeam2024dask,geopandasDevelopers2024geopandas}, with unit and smoke tests covering reassignment, purpose allocation and matrix export. The national build reported here enabled the road-mode split, child uplift, mode--time constraint and NTS0502 calibration. Every behavioural step is exposed as a command-line option, so a build for another region, year or set of assumptions differs only in configuration rather than in code; the parameters are documented in the code repository.

\section{Validation}
\label{sec:checks}

Passively generated datasets lack a ground truth to be validated against \citep{chen2016promisesBigSmallData}, and there is no independent small-area OD measurement for England and Wales against which these matrices could be checked. Accordingly, what follows is a set of consistency checks rather than a verdict on accuracy. The distinction from the diagnostics described in Section~\ref{sec:product} is one of role: the diagnostics record what each step did, while the checks below read those records back against the survey evidence and ask whether the result behaves as intended. The pipeline regenerates them on every run, so a user who rebuilds the dataset under different assumptions obtains the equivalent evidence for their own output.

\begin{figure}[htbp]
\centering
\includegraphics[width=0.92\textwidth]{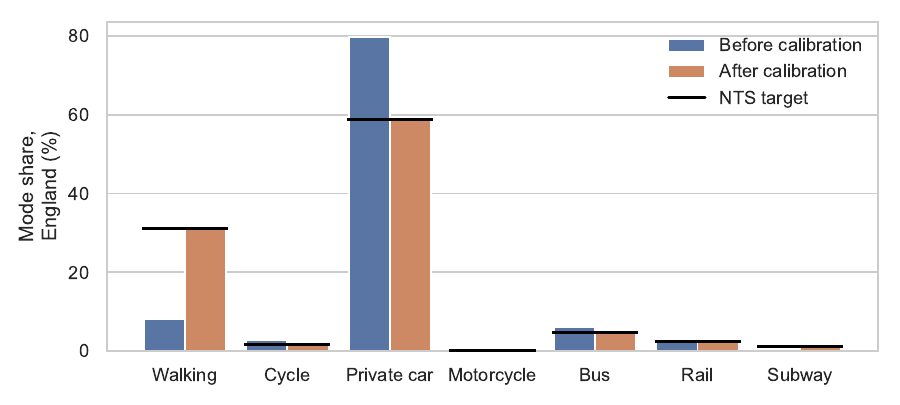}\\[0.6em]
{\small (a) Mode-share alignment with NTS targets}\\[1.2em]
\includegraphics[width=0.92\textwidth]{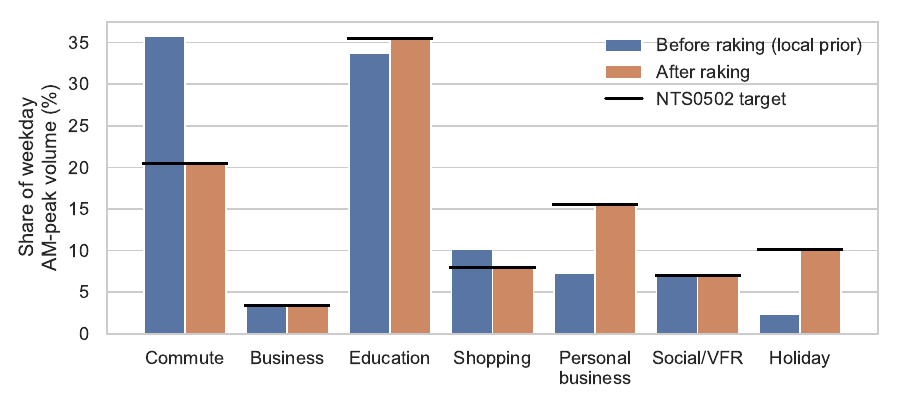}\\[0.6em]
{\small (b) Weekday AM-peak purpose shares before and after raking to NTS0502}
\caption{Consistency of the calibrated output with its survey controls. (a) Mode shares in England before and after calibration, the stage at which the typical-week product is released, against the NTS target, over the 427 origin-region and trip-length-band cells that have one; the target weights each cell's NTS shares by its uncalibrated volume. (b) National weekday AM-peak purpose shares before raking (the local prior) and after, against the NTS0502 target; raking closes every control group to within 0.01 percentage points.}
\label{fig:validation}
\end{figure}

The mode-share comparison sets the uncalibrated all-age base against the calibrated volumes across the 427 region--band--mode cells that have an NTS target: all the English cells, the remaining 25 being Welsh and untargeted. It runs on the seven output modes, with each NTS road target divided among cycle, private car, motorcycle and bus in the NTS road-split proportions, and is taken after calibration and the road-mode split, the stage at which the typical-week product is released. Across these cells the mean absolute gap falls from 5.07 to 0.01 percentage points, and from 7.86 to 0.01 when each cell is weighted by its volume. The largest movements are in the modes a mobile network is least able to distinguish: the private-car gap falls from 13.5 to 0.03 percentage points and the walking gap from 11.6 to below 0.01. In aggregate (Figure~\ref{fig:validation}a), calibration moves the base from 88.8\% road and 8.1\% walking to the NTS reference shares of 65.4\% and 31.0\%. Only one cell keeps a gap above 0.5 points: there the survey reports a mode the mobile data never recorded, and calibration cannot create travel the source does not contain. The weekday AM-peak product is further refined to the mode--time split of the trip-rate tables (Step~3), which re-targets its mode mix and so departs from these shares. Because the targets are the same evidence the calibration consumes, this check confirms that the adjustment behaves as specified; it does not establish that the matrices match travel on the ground.

Raking moves the weekday AM-peak purpose mix substantially (Figure~\ref{fig:validation}b). Before raking, the local prior places 35.7\% of AM-peak volume in commuting against an NTS0502 share of 20.4\%, and 7.3\% and 2.4\% in personal business and holiday travel against 15.6\% and 10.2\%. After raking, every control group matches its NTS0502 share to within 0.01 percentage points, while purpose composition still varies between origins with the local prior. Purpose is inferred rather than observed, so no stronger claim is available from the data alone; users whose questions turn on purpose should inspect the trip-rate priors and controls against local evidence. Every rerun regenerates the full diagnostic set, so alternative builds carry a comparable audit trail.

\section{Discussion and conclusion}
\label{sec:limitations}

Because the matrices are segmented by mode and period on a single national geography, and by purpose for the weekday morning peak, an analyst can extract exactly the flows a question needs rather than working from an undifferentiated total: AM-peak education trips for school-place and access studies, AM-peak car flows for corridor and congestion work, or bus and rail volumes for public-transport appraisal.

Five limitations bear on how the matrices should be read. First, the product is synthetic: a planning estimate, not a measurement of individual travel. Second, calibration can only act on the dimensions it compares. Mode shares are aligned within region and trip-length band, but a bias that does not surface in those margins --- a difference between the device-carrying population and the resident population, or error in the upstream inference of location and mode --- passes through into the published values. Mobile network data are known not to be representative: penetration varies between operators, and more frequent device users are both over-represented in the sample and more mobile than average, which tends to overstate mobility levels \citep{chen2016promisesBigSmallData}. Third, the NTS has covered England only since 2013, so the mode-share calibration reaches no Welsh origin: the Welsh matrices carry the observed spatial structure, the all-age uplift and the inferred purpose layer, but their mode mix is the mobile operator's classification rather than a survey-referenced one, and should be treated accordingly. Fourth, child travel is added by uplifting adult volumes at the origin, so it inherits the adult OD pattern; independent school travel and school-destination structure may be understated as a result. Fifth, purpose is inferred rather than observed, and the NTS controls are themselves survey estimates whose sampling error transfers into the product. These are properties of the underlying sources rather than defects of the workflow, and they are precisely why the pipeline is released as a configurable and auditable process rather than as a single fixed output: a user who distrusts one assumption can change it and rebuild.

Several extensions follow. The pipeline is tied to no particular vintage, so users holding a licence for a compatible feed can rebuild the matrices for another year, region or set of assumptions, and successive releases would make change over time analysable on a consistent geography. The purpose layer currently rests on population-weighted trip rates alone; incorporating destination-side evidence such as employment or retail floorspace would let purpose respond to what is at the destination rather than only to who lives at the origin. More broadly, the preserve-structure-then-calibrate workflow applies to any licensed OD source with published behavioural margins. Within the limits above, the dataset offers behavioural detail at a spatial resolution that neither the survey nor the mobile source provides alone.

\section*{Acknowledgements}

The authors thank England's Economic Heartland (EEH), BT, Network Rail and Transport for the North (TfN) for providing the data.

This joint UCL--EEH paper and accompanying dataset fall within EEH's business or operational purposes. They build on work commissioned for EEH and support its transport planning, analysis and development of modelling methods.

\section*{Data availability statement}

The published matrices and release metadata are deposited at \url{https://doi.org/10.5281/zenodo.22546252}. The \texttt{uk-travel-pipeline} source code is available under the MIT licence at \url{https://github.com/c-zhong-ucl-ac-uk/NTS-Mobile-Data-Fusion}.

The five inputs and their sources are listed in Table~\ref{tab:inputs}, where the reference against each gives its point of access and licence terms. Four are open or available to researchers under standard licences. The BT mobile network aggregate is commercially licensed and is not redistributed, so rebuilding the matrices from source requires an equivalent licence for a compatible feed. The published product is aggregate and synthetic, released at MSOA level with no individual trajectories, and a release at finer geography or time resolution would require a new disclosure assessment.

Open Government Licence v3.0 material is used with the accreditations it requires: Source: National Records of Scotland; Source: Office for National Statistics \copyright{} Crown Copyright 2024; Source: Department for Transport (2024), \emph{National Travel Survey, 2002--2023: Special Licence Access}, UK Data Service, SN: 7553.

\section*{Funding}

This research was supported by the European Research Council (ERC) under the European Union's Horizon 2020 Research and Innovation Programme (No. 949670), Proof of Concept (No. 101212775), and from the Economic and Social Research Council, UK (grant No. ES/Y010558/1).

\FloatBarrier
\bibliographystyle{plainnat}
\bibliography{references}

\end{document}